\documentclass[sigconf]{acmart}

\usepackage{booktabs}
\usepackage{graphicx}
\usepackage{subcaption}
\usepackage{adjustbox}
\usepackage{todonotes}
\usepackage{balance}
\setcopyright{none}
\renewcommand\footnotetextcopyrightpermission[1]{}
\begin{document}
\title{A Distributed and Accountable Approach to Offline Recommender Systems Evaluation}
%\titlenote{Produces the permission block, and
% copyright information}
%\subtitle{Extended Abstract}
%\subtitlenote{The full version of the author's guide is available as
% \texttt{acmart.pdf} document}

\author{Diego Monti}
\orcid{0000-0002-3821-5379}
\affiliation{Politecnico di Torino}
\email{diego.monti@polito.it}

\author{Giuseppe Rizzo}
\orcid{0000-0003-0083-813X}
\affiliation{ISMB}
\email{giuseppe.rizzo@ismb.it}

\author{Maurizio Morisio}
\affiliation{Politecnico di Torino}
\email{maurizio.morisio@polito.it}

\begin{abstract}
Different software tools have been developed with the purpose of performing offline evaluations of recommender systems. However, the results obtained with these tools may be not directly comparable because of subtle differences in the experimental protocols and metrics. Furthermore, it is difficult to analyze in the same experimental conditions several algorithms without disclosing their implementation details. For these reasons, we introduce RecLab, an open source software for evaluating recommender systems in a distributed fashion. By relying on consolidated web protocols, we created RESTful APIs for training and querying recommenders remotely. In this way, it is possible to easily integrate into the same toolkit algorithms realized with different technologies. In details, the experimenter can perform an evaluation by simply visiting a web interface provided by RecLab. The framework will then interact with all the selected recommenders and it will compute and display a comprehensive set of measures, each representing a different metric. The results of all experiments are permanently stored and publicly available in order to support accountability and comparative analyses.
\end{abstract}

\maketitle

\section{Introduction}
\label{sec:introduction}

Different authors have empirically demonstrated that offline evaluation protocols in the context of recommender systems have several weaknesses~\cite{Said2014}. For example, it is widely known that comparing results obtained in different experimental settings should be done with caution, as the slightest difference in the evaluation protocol may result in measures that are totally inconsistent~\cite{Jannach2015}.

Nevertheless, offline experiments are extremely important for comparing a large number of candidate algorithms without sustaining the costs of an online evaluation involving too many human subjects~\cite{Gunawardana2015}. After having pruned the set of available systems, it is however advisable to analyze them in a real environment for obtaining more conclusive results.

In this paper, we propose a way of overcoming the problem of comparing offline evaluation results obtained from different recommendation algorithms in heterogeneous settings. To this end, we designed and implemented an open source evaluation framework for top-$k$ prediction methods, called RecLab,\footnote{\url{https://github.com/D2KLab/reclab}} that is capable of interacting with several recommenders using RESTful APIs.

The responsibilities of the evaluator and the recommender are clearly separated because of the distributed architecture of the system. The evaluator is in charge of building a training set, selecting a set of users to whom recommend the items, and computing the evaluation metrics. On the other hand, the recommender must be capable of predicting a list of the most appropriate items for each user, given the information available in the training set.

The configuration parameters of each experiment are left to the user of the toolkit, who is free to choose the dataset, the splitting strategy, the size of the test set, the length $k$ of the recommended lists, and the rating threshold between relevant and irrelevant items. The experiment can be designed and run by simply interacting with a web-based GUI provided by the toolkit. Other researchers can easily plug their recommender systems into the evaluation pipeline by implementing the APIs defined by RecLab and by deploying them on a server. Thanks to this approach, it is possible to compare, in a controlled environment, different algorithms and techniques without necessary disclosing their implementation details. The results of each experiment, along with the respective configuration parameters, are publicly available to support accountability and comparative analyses of the results.

The remainder of this paper is structured as follows: in Section~\ref{sec:related} we review related works. In Section~\ref{sec:framework} we introduce the main design choices behind our evaluation framework, while in Section~\ref{sec:interaction} we describe how different recommenders can interact with the evaluator. In Section~\ref{sec:metrics} we explain the mathematical details of the metrics computed during the evaluation phase. We present and discuss our results in Section~\ref{sec:results} and, in Section~\ref{sec:conclusion}, we provide the conclusions.
\section{Related Work}
\label{sec:related}

To the best of our knowledge, the problem of evaluating a recommender system was first addressed by Herlocker \textit{et al.}~\cite{Herlocker2004}. In their work, the authors argue that an offline experiment, in order to be complete and trustworthy, should rely on a comprehensive set of metrics. They review several indicators usually exploited by different researchers and they classify them into three categories: predictive accuracy metrics, classification accuracy metrics, and rank accuracy metrics. RecLab is mostly based on rank accuracy metrics, as those are usually employed in real scenarios.

Jannach \textit{et al.}~\cite{Jannach2015} compared several recommendation algorithms in an offline experiment considering different splitting protocols and metrics. Their results suggest that some algorithms, despite their high accuracy, tend to only recommend popular items that are probably not very interesting to users. For this reason, it is not advisable to evaluate algorithms by relying only on accuracy metrics. With these conclusions in mind, we decided to include in RecLab a comprehensive set of seven metrics.

Said and Bellog{\'{\i}}n~\cite{Said2014} analyzed several recommender systems evaluation frameworks in order to check if their results are consistent. They discovered that the values obtained with the same dataset and algorithm may vary significantly among different toolkits. These discrepancies are mainly caused by the data splitting protocol, the strategy used to generate the candidate items, and the implementation details of the evaluation metrics. In order to support the reproducibility of the results, we also store the experimental settings together with the measurement outcomes.

In several domains, researchers have already realized comprehensive benchmarking platforms for evaluating different algorithms in a controlled environment relying on a standardized set of metrics. Texygen~\cite{Texygen2018} is a toolkit for measuring the performance of text generation models; however, differently from RecLab, it does not include a web-based interface. On the other hand, GERBIL~\cite{Gerbil2015} is an evaluation framework for semantic entity annotation that permanently stores and publishes on the Web the results of each experiment for reproducibility purposes. 

\section{Evaluation Framework}
\label{sec:framework}

RecLab has been implemented as a distributed web application: its users can setup the experimental environment by simply visiting a web page. This step is crucial for the correct execution of the measurements, as selecting inconsistent or wrong values may lead to results that are extremely difficult to interpret~\cite{Jannach2015}.

In details, the experimenter needs to specify the following parameters before starting an evaluation:

\begin{itemize}
    \item the initial rating dataset;
    \item the technique used to split the dataset;
    \item the size of the training and the test set;
    \item the length $k$ of the lists of recommended items;
    \item the threshold between negative and positive ratings;
    \item the list of recommenders to be evaluated.
\end{itemize}

We included in this evaluation framework three widely used rating datasets: MovieLens 100K,\footnote{\url{https://grouplens.org/datasets/movielens/100k/}} MovieLens 1M,\footnote{\url{https://grouplens.org/datasets/movielens/1m/}} and HetRec LastFM.\footnote{\url{https://grouplens.org/datasets/hetrec-2011/}} The MovieLens datasets are among the most popular recommender systems datasets about movies preferences~\cite{Harper2015}, while the last one is particularly interesting as it contains the number of times each user listened to a specific artist on LastFM~\cite{Cantador2011}. Other datasets can be easily added by editing a configuration file.

We provide two different methods for splitting the rating dataset $\mathcal{R}$ in a training set $\mathcal{R}_{train}$ and a test set $\mathcal{R}_{test}$ such that $\mathcal{R} = \mathcal{R}_{train} \cup \mathcal{R}_{test}$. The first one is a random splitting method that assigns each rating $\rho \in \mathcal{R}$ to the test set or the training set according to a probability specified by the user, that is proportional to the expected size of the test set. In general, this method should be the preferred one when no temporal information is available~\cite{Gunawardana2015}; the default size of the test set is the 20\% of the dataset.

A second splitting technique is based on the timestamps associated to the ratings: the whole dataset is ordered from the oldest to the newest rating; then, the first ones are assigned to $\mathcal{R}_{train}$, while the others to $\mathcal{R}_{test}$. This protocol simulates the behaviour of a recommender introduced at a certain point in time in the system. While the HetRec LastFM dataset does not include any timestamp, the MovieLens ones do. However, their values probably do not represent when users watched a certain movie, but when they rated it on the platform.

Another fundamental parameter of the experiment is the length $k$ of the lists of recommended items, as it will deeply influence all the metrics computed by the evaluator. This value should be set according to the number of items that the final application will display to its users. Typical values for this criterion are $5$ or $10$.

The experimenter also needs to specify what is the threshold between negative and positive ratings: only ratings with a value strictly greater than the threshold will be considered \emph{likes} during the evaluation phase. Many recommenders will only analyze positive ratings during the training phase. However, this is beyond the scope of the evaluator, so it is a responsibility of the recommender to properly load the ratings. The most appropriate value for this parameter depends on the dataset: a typical setting for MovieLens datasets is $3$, thus only $4$ and $5$ stars ratings are considered positive. For the HetRec LastFM dataset, as the rating value represents the number of times a user listened to an artist, any small number may be reasonable, including $0$.

For demonstrative purposes, we included in RecLab a set of recommender systems that follow the interaction protocol described in Section~\ref{sec:interaction}. However, anyone is encouraged to implement other techniques for the purpose of evaluating them with this framework. Further recommenders can be added by simply inserting their URIs in a configuration file present in our repository. All available recommenders are then displayed to the experimenter, for letting her select which ones to evaluate.

In details, we have realized the classical most popular and random recommenders. Our most popular recommender is \emph{personalized}: it will never suggest to a certain user items already rated by the same user in the training set. On the other hand, the random recommender will select any item available in the training set with an equal probability.

Furthermore, we have included the MyMediaLite~\cite{Gantner2011} implementations of the Item KNN, User KNN, BPRMF, and WRMF recommender systems using their default settings.\footnote{\url{http://www.mymedialite.net/documentation/item_prediction.html}} BPRMF is a recommendation algorithm based on a Bayesian ranking optimization method~\cite{Rendle2009}, while WRMF is weighted matrix factorization technique~\cite{Hu2008}.

\section{Interaction Protocol}
\label{sec:interaction}

\begin{figure}
\includegraphics[width=.65\linewidth]{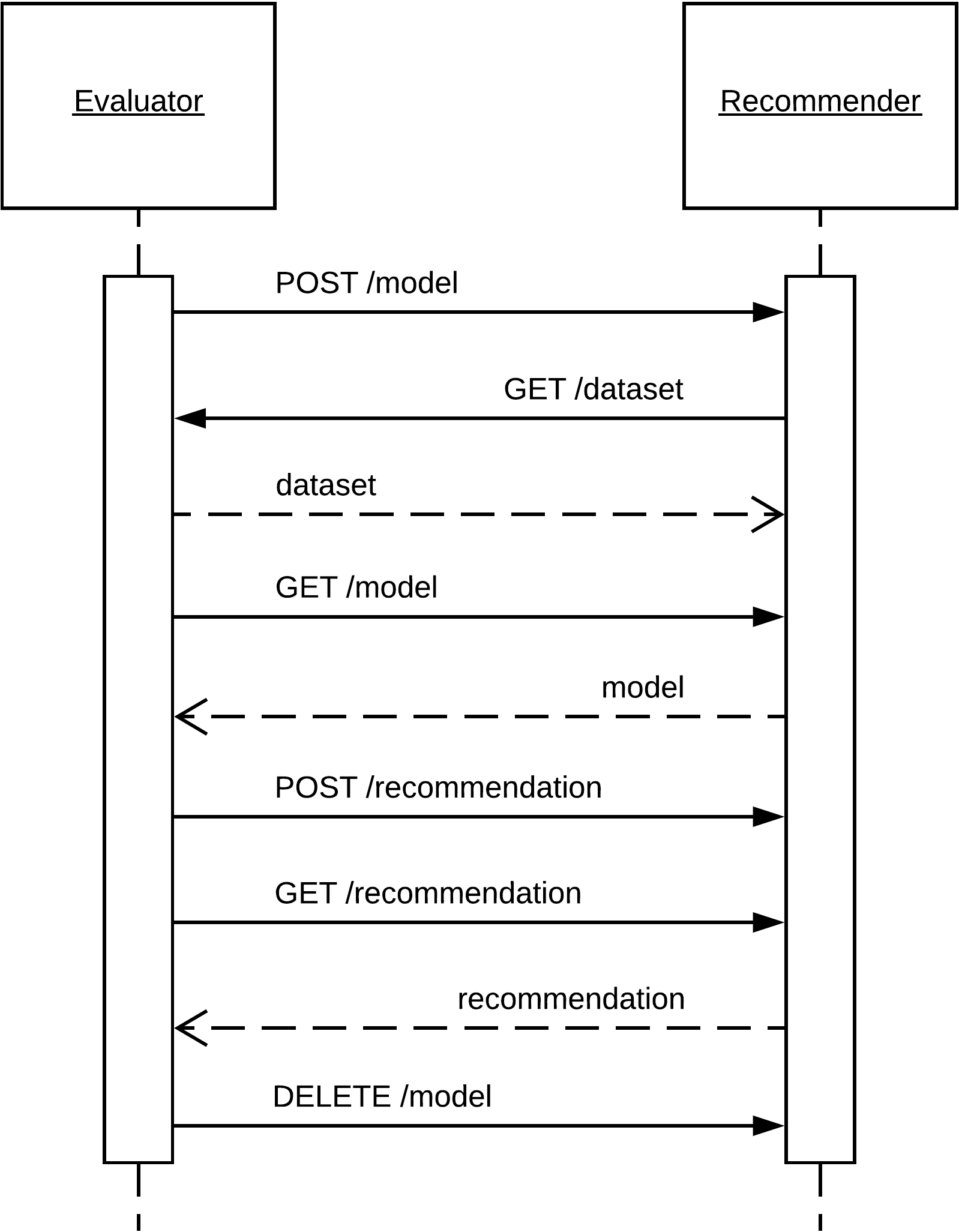}
\caption{Simplified UML sequence diagram}
\label{fig:sequence}
\end{figure}

RecLab is a distributed evaluation framework. For this reason, it exploits consolidated web standards to create a communication channel among itself and the recommenders under analysis: the overall protocol is graphically depicted as a UML sequence diagram in Figure~\ref{fig:sequence}. This interaction is initiated when the experimenter decides to execute a new evaluation, and it is repeated for every recommender selected as part of it.

First, the evaluator requests the recommender to train a new recommendation model with a \textsc{post} on the resource \texttt{/model}. It provides to the recommender a URI from which it can download the training set and the rating threshold, as specified by the experimenter. Only the ratings with a value strictly greater than the threshold should be considered as positive feedbacks.

The recommender can now retrieve the training set from the provided URI. Each training set is specific for a particular experiment, but it is created at run time by the evaluator using the configuration settings specified by the user. The training set consists of a list of ratings, where each rating associates a user, an item, and, optionally, a timestamp to a numerical value.

Now the recommender has all the information required to perform the training process. Meanwhile, the evaluator will start asking asynchronously to the recommender if this phase has ended with a \textsc{get} of \texttt{/model}. When the recommender is ready to suggest items, the evaluator is informed of that and it can proceed to the next step.

The evaluator asks the recommender, with a \textsc{post} on the resource \texttt{/recommendation}, to create a list of $k$ items for each user specified in the payload of the request. The list of users to whom recommend the items contains all the users available in the test set, while the value of $k$ was initially provided by the experimenter. Note that it is a responsibility of the recommender avoiding to suggest items already rated by a particular user in the training set. In general, there is no guarantee that the test set will only contain users and items available in the training set.

Because also the recommendation phase may be time consuming, it is considered asynchronous, similarly to the training one. The evaluator starts asking with a \textsc{get} on the same resource if the lists of recommendations are ready. When they are correctly retrieved, the evaluator asks the recommender to \textsc{delete} the \texttt{/model} to avoid consuming memory, while it begins to compute the evaluation metrics detailed in Section~\ref{sec:metrics}.

\section{Metrics}
\label{sec:metrics}

In order to better analyze the recommender systems under evaluation from different perspectives, we decided to include in RecLab a comprehensive set of seven different metrics. In fact, it is not possible to accurately evaluate in an offline experiment a set of recommenders by only relying on a single indicator~\cite{Herlocker2004}. We selected not only classic metrics like coverage and precision but also less widespread ones like novelty, diversity, and serendipity.

In the following, we define $\mathcal{U}$ as the set users $\upsilon \in \mathcal{U}$, $\mathcal{I}$ as the set of items $\iota \in \mathcal{I}$, and $\mathcal{R}$ as the set of ratings $\rho \in \mathcal{R}$.

Furthermore, we define $rec(\upsilon, k)$ as the list of the top-$k$ items recommended to user $\upsilon$ and $ref(\upsilon)$ as the set of items rated positively by user $\upsilon$ in the test set $\mathcal{R}_{test}$.

\begin{description}
\item[Coverage]
The coverage of a recommender is a measure that represents the number of items in the catalog over which the system can make suggestions~\cite{Gunawardana2015}. Given the lists of recommended items for each user in the test set, we compute the percentage of distinct suggested items with respect to the distinct items available in the training set.

\begin{equation*}
\mathrm{coverage}(k) = \frac{|\bigcup_{\upsilon \in \mathcal{U}_{test}} \mathcal{I}_{\mathrm{rec}(\upsilon, k)}|}{|\mathcal{I}_{train}|}
\end{equation*}

This metric captures if the recommender is capable of suggesting enough various items to each user, or if it always proposes the same items to all the users. Coverage should be analyzed together with precision, otherwise it is clear that random recommendations will achieve optimal results~\cite{Herlocker2004}.

\item[Precision]
Precision, in the context of information retrieval, represents the fraction of selected documents that are relevant. For a recommender system, it measures the fraction of recommended items that are liked by a user~\cite{Sarwar2000}.

\begin{equation*}
\mathrm{precision}(k) = \frac{1}{|\mathcal{U}_{test}|} \cdot \sum_{\upsilon \in \mathcal{U}_{test}} \frac{|rec(\upsilon, k) \cap ref(\upsilon)|}{k}
\end{equation*}

In order to avoid overestimating the value of precision, we assume that all non-rated items are irrelevant~\cite{Steck2013}.

\item[Recall]
Complementary to precision, recall represents the fraction of relevant documents that have been selected. In the context of recommender systems, it measures the fraction of correctly recommended items with respect to all the items the are liked by a user~\cite{Sarwar2000}.

\begin{equation*}
\mathrm{recall}(k) = \frac{1}{|\mathcal{U}_{test}|} \cdot \sum_{\upsilon \in \mathcal{U}_{test}} \frac{|rec(\upsilon, k) \cap ref(\upsilon)|}{|ref(\upsilon)|}
\end{equation*}

If the set of items rated positively by a user is empty, we assume that the recall for that user is $0$ by definition.

\item[NDCG]
The Normalized Discounted Cumulative Gain is another information retrieval metric, that also considers a logarithmic gain related to the position of each correctly predicted item~\cite{Jaervelin2002}. This metric reveals if a recommender is capable of correctly predicting items at the top of the list.

\begin{equation*}
\mathrm{dcg}(k) = \frac{1}{|\mathcal{U}_{test}|} \cdot \sum_{\upsilon \in \mathcal{U}_{test}} \sum_{i = 1}^{k} \frac{|\{\iota_i\} \cap ref(\upsilon)|}{log_2(i + 1)}
\end{equation*}

Where $\iota \in rec(\upsilon, k)$. The DCG value needs to be divided by the ideal DCG for normalization. The ideal DCG can be computed with the same formula, assuming that all recommended items are relevant for the associated user.

\item[Novelty]
This metric rewards algorithms capable of suggesting items that belong to the long-tail of the catalog, and so it is unlikely that they are already known by a certain user~\cite{Vargas2011}.

In this way, it is possible to check that the recommended items are not too popular and obvious.

\begin{equation*}
\mathrm{novelty}(k) = - \frac{1}{|\mathcal{U}_{test}| \times k} \cdot \sum_{\upsilon \in \mathcal{U}_{test}} \sum_{i = 1}^{k} \log_2 \mathrm{freq}(\iota_i)
\end{equation*}

Where $\iota \in rec(\upsilon, k)$ and $freq : \mathcal{I} \to [0, 1]$ represents the probability of observing the item $\iota$ in $\mathcal{I}_{train}$. We also assume that $\log_2(0) \doteq 0$ by definition.

\item[Diversity]
The metric of diversity is inspired by the metric of Intra-List Similarity proposed by Ziegler \textit{et al.}~\cite{Ziegler2005}. It measures how much the items included in the recommended lists are diverse. A higher diversity may be beneficial for the users, as they are encouraged to better explore the catalog~\cite{Noia2014}.

\begin{equation*}
\mathrm{diversity}(k) = \frac{1}{|\mathcal{U}_{test}|} \cdot \sum_{\upsilon \in \mathcal{U}_{test}} \frac{\sum_{\forall{i}, \forall{j} : 0 < i < j}^k 1 - \mathrm{sim}(\iota_i, \iota_j)}{k \times (k - 1)}
\end{equation*}

Where $\iota \in rec(\upsilon, k)$ and $sim : \mathcal{I} \times \mathcal{I} \to [-1, 1]$ is a similarity measure between two items. We decided to exploit the cosine similarity computed between the vectors representing the users who liked the two items in the training set.

The resulting value is a number in the interval $[0, 2]$: higher values imply an higher diversity.

\item[Serendipity]
Serendipity can be defined as the capability of identifying items that are both attractive and unexpected~\cite{Gemmis2015}. It is possible to measure the serendipity of a recommender by computing its precision after having discarded the items suggested by a \textit{primitive} recommender~\cite{Ge2010}.

\begin{equation*}
\mathrm{serendipity}(k) = \frac{1}{|\mathcal{U}_{test}|} \cdot \sum_{\upsilon \in \mathcal{U}_{test}} \frac{|(rec(\upsilon, k) \setminus prim(k)) \cap ref(\upsilon)|}{k}
\end{equation*}

Where $prim(k)$ is the set of the top-$k$ most popular items available in the training set. We can, in fact, suppose that popular items are already known by several users, and thus they cannot contribute to the serendipity of the suggestions.

\end{description}

This set is not final, as RecLab can be easily expanded in order to compute additional metrics that the community considers useful.

\section{Experimental Results}
\label{sec:results}

\begin{table*}
\begin{tabular}{@{}llllllll@{}}
\toprule
Algorithm    & Coverage & Precision & Recall   & NDCG     & Novelty  & Diversity & Serendipity \\ \midrule
Random       & 1.000000 & 0.005152  & 0.002526 & 0.005069 & 13.37526 & 0.966485  & 0.005003    \\
Most Popular & 0.017920 & 0.145146  & 0.084294 & 0.158512 & 8.580345 & 0.600524  & 0.071869    \\
Item KNN     & 0.473527 & 0.212028  & 0.137608 & 0.224146 & 10.55504 & 0.788987  & 0.196637    \\
User KNN     & 0.141732 & 0.263337  & 0.189679 & 0.295034 & 9.052157 & 0.657436  & 0.205550    \\
BPRMF        & 0.326907 & 0.225464  & 0.148515 & 0.247625 & 9.473122 & 0.717023  & 0.176972    \\
WRMF         & 0.120554 & 0.258400  & 0.169925 & 0.287808 & 9.138422 & 0.667673  & 0.210835    \\ \bottomrule
\end{tabular}
\caption{Evaluation results with the MovieLens 1M dataset and a random splitting}
\label{tab:movielens-random}
\end{table*}

\begin{table*}
\begin{tabular}{@{}llllllll@{}}
\toprule
Algorithm    & Coverage & Precision & Recall   & NDCG     & Novelty  & Diversity & Serendipity \\ \midrule
Random       & 0.993719 & 0.017555  & 0.002910 & 0.017394 & 13.41860 & 0.963699  & 0.016938    \\
Most Popular & 0.037411 & 0.257487  & 0.053148 & 0.273653 & 8.546251 & 0.528352  & 0.066517    \\
Item KNN     & 0.344894 & 0.231296  & 0.056491 & 0.244158 & 9.759138 & 0.670575  & 0.095962    \\
User KNN     & 0.117422 & 0.275042  & 0.062094 & 0.293720 & 8.842892 & 0.567265  & 0.114470    \\
BPRMF        & 0.220918 & 0.265339  & 0.062845 & 0.282460 & 9.083545 & 0.611382  & 0.112675    \\
WRMF         & 0.136537 & 0.276164  & 0.065600 & 0.297178 & 8.941997 & 0.587175  & 0.121929    \\ \bottomrule
\end{tabular}
\caption{Evaluation results with the MovieLens 1M dataset and the timestamp splitting}
\label{tab:movielens-timestamp}
\end{table*}

\begin{table*}
\begin{tabular}{@{}llllllll@{}}
\toprule
Algorithm    & Coverage & Precision & Recall   & NDCG     & Novelty  & Diversity & Serendipity \\ \midrule
Random       & 0.708420 & 0.000584  & 0.000632 & 0.000623 & 15.30801 & 0.998417  & 0.000584    \\
Most Popular & 0.001692 & 0.066773  & 0.069242 & 0.076932 & 7.736651 & 0.632728  & 0.019161    \\
Item KNN     & 0.235489 & 0.126168  & 0.131249 & 0.143234 & 12.56312 & 0.774870  & 0.101486    \\
User KNN     & 0.031299 & 0.158652  & 0.164218 & 0.193206 & 8.735683 & 0.717815  & 0.115711    \\
BPRMF        & 0.024597 & 0.078715  & 0.081908 & 0.086609 & 8.280671 & 0.752627  & 0.038800    \\
WRMF         & 0.015617 & 0.164809  & 0.170302 & 0.201023 & 8.849644 & 0.763729  & 0.123992    \\ \bottomrule
\end{tabular}
\caption{Evaluation results with the HetRec LastFM dataset and a random splitting}
\label{tab:lastfm}
\end{table*}

To prove the effectiveness of RecLab, we performed three different experiments with the recommender systems described in Section~\ref{sec:framework} whose implementation details are available in our repository.

In the first one, we selected the MovieLens 1M dataset and we chose a random splitting protocol. For the other parameters, we used the default values of the framework: the test set size is the 20\% of the dataset, the length $k$ of the recommended lists is equal to $10$, while the rating threshold is equal to $3$. The results of this first experiment are reported in Table~\ref{tab:movielens-random}.

In the second experiment, we only changed the splitting protocol to the timestamp-based one, while we retained all other parameters unmodified. The results are reported in Table~\ref{tab:movielens-timestamp}.

Finally, for performing the third experiment, we selected the HetRec LastFM dataset. All other parameters are the same of the first experiment, but the rating threshold, which is now equal to $0$. The results of this last experiment are reported in Table~\ref{tab:lastfm}.

As expected, the random recommender always achieves the best coverage, novelty, and diversity, while also obtaining the worst precision, recall, NDCG, and serendipity. On the other hand, the most popular recommender has a very low coverage and novelty, but it also has interesting values of precision and NDCG, especially with the MovieLens dataset. Note its impressive increase in terms of precision obtained by simply changing the splitting protocol.

The measures of serendipity are not exactly zero because this implementation of the most popular recommender suggests a \emph{personalized} list of popular items, avoiding the ones already rated by the same user in the training set.

If we ignore the random suggestions, the Item KNN recommender obtains the best results in terms of coverage, novelty, and diversity in all the experiments. Regarding the metric of precision, we observe interesting results with the User KNN recommender in the first experiment, with the User KNN and the WRMF in the second one, and with the WRMF in the last one. We observe a dramatic decrease in precision of the BPRMF recommender in the LastFM experiment, probably due to the characteristics of the dataset.

In the last experiment, the WRMF algorithm achieves the best values of precision, recall, NDCG, and serendipity. However, it also scores a very low coverage; in contrast, the Item KNN recommender has a fair precision and an interesting coverage. For this reason, it would be useful to compare these algorithms in an online experiment involving human subjects.
\section{Conclusion and Future Work}
\label{sec:conclusion}

In this paper, we have introduced RecLab, an open source framework for evaluating top-$k$ recommender systems in a distributed setting. The main aim of this work is to support the accountability and the reproducibility of the results of the experiments by permanently storing and publicly displaying their configuration parameters and numerical outcomes.

RecLab is based on a RESTful interaction protocol that enables researchers to evaluate different recommenders created with heterogeneous technologies in a common experimental context and with a comprehensive set of metrics.

The results of each experiment can be easily retrieved and automatically processed using a machine-readable format.

We exploited RecLab for performing three experiments involving all the recommenders at our disposal. We empirically validated the importance of considering different metrics in order to execute a reliable evaluation and we observed the impact of the configuration parameters on the outcome of the experiments.

As future works, we plan to integrate more rating datasets and other recommendation techniques. We also envision the possibility of enhancing the interaction protocol in order to let the experimenter specify the configuration parameters of each recommender.

%\balance
\bibliographystyle{ACM-Reference-Format}
\bibliography{references} 

\end{document}